# The government of state's power bodies by means of the Internet

Bercea L., Nemțoi G., Ungureanu C.

**Abstract**— The electronic government involves developing the informational society, which refers to an economy and a society in which the access, acquisition, memorizing, taking, transmitting, spreading and using the knowledge accede to a decisive role. The informational society involves changes in the domains of administration (e-Government), business (electronic commerce and e-business), education (long distance education), culture (multimedia centers and virtual libraries), mass- media (TV, video advertising panels), and in the labor manner (tele-work and virtual commuting).The e-government refers to the interaction between the Government, Parliament and other public institutions with the citizens by the electronic means.

——————————— ◆ ———————————

## 1 INTRODUCTION

The informing the citizen of the law projects that are debated and allowing them to express their opinion, tax payment by the taxpayers and complaints form completion, all of these made online are effective means the state makes available in order to exert the fundamental rights of the citizens. In this way, the government is made in a vertical position, from top to bottom- from the state to the citizen. Against this concept, the interaction between the citizen and the state must be assimilated, and that is from bottom to top- the electronic democracy, by which the citizen can use an electronic system in order to communicate with other citizens or even with the institutions of the public authority. The electronic democracy is organized for the citizens and by the state, to make possible evaluating the improvement proposals for the government process and debating the governmental activities or actions. All these problems are made on the internet, by the personal opinions sites, dedicated portal forums, chats and discussions list. The electronic democracy came up as a necessity, by which the state transfers information regarding the efficient public services to the citizens, to perfect and modernize the democratic system. In this way, the e-government is a complex phenomenon that refers to multiple connections that are between the authorities and the persons or companies that exist in a society.

————————————————

• *Bercea L. is with the Faculty of Law and Administrative Sciences, West University of Timisoara, Romania*

• *Nemtoi G. is with the Law and Judicial Sciences Department, Faculty of Economics and Public Administration, Stefan cel Mare University of Suceava, 13 University Str., Suceava, Romania*

• *Ungureanu C. is with the Faculty of Economics and Public Administration, Stefan cel Mare University of Suceava, 13 University Str., Suceava, Romania*

## 2 MATERIALS AND METHODS – THE NEED FOR ELECTRONIC DEMOCRACY

All over the world, the governments look with a great interest to the new technologies and future electronic services, but their vision stops when taking into calculation the technical aspects. Although the new technologies are the ones that allow the implementation of electronic government, it must be said that an electronic implementation of public services means major changes, on an internal but also external level of the institution, or in short it means a new approach to serving the citizen. The e-government has gone through 3 phases: forms presentation, the possibility to download the necessary forms to interact with the authorities and the possibility to complete the forms on-line and make transactions, implicitly paying the debts to the public authorities.

The development degree of the electronic government and democracy depends directly on the access of the population to the internet. In order to ease the citizen's access to the electronic services, the web sites of the public institutions must present upgraded information in a clear and concise. The information must be easily accessible with any browser and be presented both in an international and a national language. The e-government represents the process to recreate the public sector by digitalization and new management techniques of the information. By this process are followed the increase of the citizen's political involvement and the efficiency of the administrative apparatus. Seen from this point of view, the e-governance reunites three situations as base models of the system and those are:

- The technical paradigm, that consists of the use of the electronic technologies in communication (e-mail, chat, internet sites).
- The managerial paradigm, that aims to apply new management modes for the information;
- The functional paradigm, that refers to the quantum of the active involvement of the population by accessing the internet for the administrative apparatus's services.

Within the government process, three categories of participants are involved:
- The public participants, which represent the public institutions of the central and local administration.
- The given country's participants - the citizens.
- The participants of the private foreign companies that organize commercial activities.

These participants create specific communicational and transactional relations, that trigger existent intern flows within a participants class (the relations between the Parliament and Government), or external flows between two classes of participants (citizen and public institution). The main components of the e-government will be described below.

## 3 THE OBJECTIVES OF THE E-GOVERNMENT COMPONENTS

Starting from the premises that, by the objectives of the informational society, the efficiency of the administrative act is desired, in 2003, in Romania, the National Electronic System (SEN) launches at the www.e-guvernare.ro address an informational system of public utility which has, as a main objective, assuring the access to the public information and supplying public services. The SEN represents the main access point to information and electronic services of the public administration, which offers to the users a wide range of electronic services destined to the private companies, access to forms, access to administrative procedures of the public administration etc. In present times, the SEN has the following services implemented:
- The main SEN portal and access for the interested to information regarding contact points of the public institutions (http://www.e-guvernare.ro).
- Information regarding the processing procedures.
- Access to electronic services of the central and local public administration (https://formularunic.e-guvernare.ro).

The key objectives of the e-government portal are:
- The public electronic services must be oriented towards the user.
- The electronic services must be secured and trustworthy.
- The services must be supplied unitarily and through a simple, intuitive and standardized interface.
- Transparence in the public administration.

In the development process of the electronic government, the SEN role is to assure infrastructure services for the users:
- Secured routing engine of the messages for the electronic government services.
- Unique access point to the public services supplied by electronic means.
- Single Sing-On for the electronic government services.
- Transaction engine for the electronic government services.
- Messaging engine for the electronic government services.
- Reporting engine for the electronic government services.
- A collection of services' documents, functioning libraries that are at the disposal of any independent developer wishing to create applications with the SEN.

A fact must be taken into account- that the whole communication in the transactional environment between the users must be done by ISO, PDF, xml documents. The www.e-guvernare.ro portal groups the four components of the IT system:
- G2G-Government to Government

The G2G is an IT system by which can be followed the communication by electronic means between the public institutions and consist in collecting data contained by the public institutions and applying solutions for their solving, the G2G means communication between several public institutions to solve an unique problem of the citizen.

Integrating the intergovernmental systems is the first step in implementing electronic government solutions.
- G2E-Government to Employee

The G2E involves the existence of the online management of the relation between the government and employees by the TIC. This system involves 2 types of applications: applications that ease the fulfilling of work tasks and applications for the management of the employee situation of the government employee.
- G2C-Government to Citizens

The G2C is a component of the e-government portal which implements the relations between the government and the citizen. In essence, the G2C represents the most important component because the relation between the citizen and government is interactive and is made by three application types: applications that make the major function of the state to facilitate transparency and free change of information between the citizen and the state, application that consist in fulfilling the state's role as an electronic governmental services supplier and applications that make the electronic vote. The G2C component is decisive in the effective participation of the citizen to the state's governmental political life.
- G2B-Government to Business

The relations between the companies and the state are an important cause in developing this component because the private companies represent the economical growth engine of the state, therefore the relation supplied by the administrative authorities to the companies have, as a purpose, creating an informational area that is to simplify the interaction between the state and companies

There are two types of applications: *the internet public acquisitions systems* and *the services offered by the state to the private sector*. Out of these applications, the most useful are the one that require less time and money for the government and for the companies.

## 4 RESULTS AND DISCUSSION

The e-government is a modern concept, which might be extremely useful for each of us. The part of e-government solutions is described by the site having the same name and assumes "the reducing of bureaucracy administrative borders, as well as facilitating the access to public information and services". In this way, by means of the electronic government, the public services become more efficient, more transparent and more accessible, no matter the time or place. The unique point of access to these services and public information, established according to Law 161/2003 is The National Electronic system (NES) (source: www.e-guvernare.ro).

The main objectives of EU within e-administration, which have to be reached in Romania also, as result of the agreements signed, refer in generally to:
- making efficient the public administration activities;
- Facilitating the access of citizens to various range of electronic services;
- A higher transparency of the decisional act over the administrative instrument; all these would generate the mentioned reduction over bureaucracy costs.

According to European agenda established two years ago, this objective has to be reached until 2010, by all member states of EU. This fact, regarded at the current level of informatics over public administration of Romania, does not seem a demarche too easily to accomplish. Even if the funds allotted in order to gather the know-how over implementing the e-government technologies were enough, according to specialists, one might see a deficiency at the level of understanding these notions, at the administrative level, which translates into a preponderant orientation towards citizens, in the detriment of business environment, according to the studies carried out. The result of implementing the information system is a main reason on depletion of the bureaucracy frame.

## 4 CONCLUSIONS

The electronic government is one of the important manifestations of the contemporary society, because it involves institutions of the state, public and private organizations and especially the citizen's activity, him participating electronically to the government activity. Although considered as a problematic positively accepted by the civil society, the electronic government, besides its benefits involves a series of risks such as: the reaction of the employees within the implementing organization, the neutralization risk by the beneficiaries because of the lack of internet access or fears connected to the security of the spread information, the technological process that could determine the need to continuously upgrade the system etc. In conclusion, implementing this portal means a new development stage to the state institutions and, finally, an evolution in de-concentrating the administrative bureaucracy.

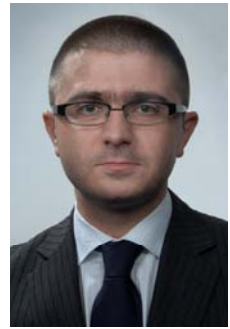

Lucian Bercea – Vice-Dean, Associate professor, Faculty of Law and Administrative Sciences, West University of Timisoara, Romania. PhD - *Magna cum laude*, Master in Law, Business Law. Author of 5 books published in Commercial Law at international publications, amongst *On why law should nowadays interfere with biology*, în F. Tremmel, B. Meszaros, C. Fenyvesi (ed.), Studia Iuridica Auctoritate Universitatis Pecs Publicata: Orvosok és jogászok a büntető igazságszolgáltatásban. Dezsö László Emlékkönyv, University of Pecs, Hungary; *Consumer credit: a comparison between the right of withdrawal from the credit agreement and the right of early repayment of the credit*, in *Diritto e Politiche dell'Unione Europea*, G. Giapichelli Editore, Italy, no. 1/2009, 2 papers published at international journals, 25 papers in publications of specialty, as Journal of Commercial Law, 25 papers at international conferences, 10 participations at national conferences.

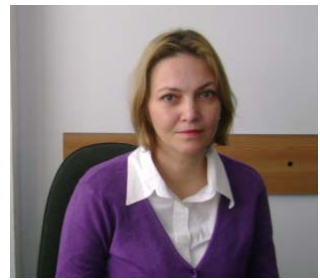

Nemțoi Gabriela - Lecturer, PhD candidate, Faculty or Economics and Public Administration, Department of Law and Judicial Sciences, 'Stefan cel Mare' University of Suceava. Author of 30 papers, participant to 5 international conferences, with papers included to journals, 4 scientific papers published in data basis and EBSCO included, PhD Candidate at State Academy of Republic of Moldova. The PhD thesis title: "The part of people's sovereignty over state organization of capacities".

Ungureanu Ciprian – Lecturer, PhD candidate, Faculty or Economics and Public Administration, Department of Public Administration, 'Stefan cel Mare' University of Suceava. PhD Candidate at State Academy of Republic of Moldova, Law section, author of 5 papers *B/B+*, participant at 5 international conferences, member within 10 grants, significant scientific papers - From Red Tape to Smart Tape. –NISPA granted project.